WHY RADIO QUIET QUASARS ARE PREFERRED OVER RADIO LOUD QUASARS REGARDLESS OF ENVIRONMENT AND REDSHIFT

David Garofalo[1], Max North[2], Leanne Belga[1], Kenzi Waddell[1]

1. Department of Physics, Kennesaw State University
2. Department of Information Systems, Kennesaw State University

Abstract

Evidence has accumulated suggesting the clustering of radio loud quasars is greater than for radio quiet quasars. We interpret these results in a context in which the fraction of radio loud quasar formation is $f_{RLQ} \leq f_{RQQ}$ compared to that for radio quiet quasars for all environments and redshift. Because we assume that post-merger cold gas onto large black holes produces either a radio loud or a radio quiet quasar, we show that for largest black hole masses that live in largest dark matter halos, $f_{RLQ}$ approaches 0.5 from below but does not exceed it, such that in rich clusters the formation of a radio loud quasar tends to be equally likely to occur as a radio quiet quasar. In dark matter halos with smaller mass, by contrast, radio quiet quasars are more likely to form and the likelihood increases inversely with dark matter halo mass. As a result, averaging over a population of radio loud and radio quiet quasars will necessarily generate lower average black hole masses for the radio quiet subgroup. Hence, despite the fact that the formation of radio quiet quasars is preferred over radio loud quasars in any environment, at any mass scale, at any luminosity, or redshift, averaging over a range of radio loud quasars will give the appearance they are preferred in cluster environments over radio quiet quasars. We show how this also accounts for the order of magnitude difference in the total number of jetted active galaxies compared to non-jetted counterparts.

I. Introduction

Exploration of the richness factor for radio loud and radio quiet quasars shows the former to be more clustered (Worpel et al 2013; Retana-Montenegro & Rottgering 2017). Retana-Montenegro & Rottgering (2017) have explored quasar clustering for 45441 and 3493 radio quiet (RQQ) and radio loud quasars (RLQ), respectively, from the SDSS and FIRST surveys, finding that RLQ inhabit more massive halos compared to RQQ. At first sight, these results seem compatible with recent work on merger signatures in these AGN subgroups which find RLQ associated with merger signatures at a higher rate compared to RQQ (Ivison et al 2012; Wylezalek et al 2013; Chiaberge et al 2015; Hilbert et al 2016; Noirot et al 2018; Zakamska et al 2019). If quasars are formed by cold gas funneled into galactic nuclei (Barnes & Hernquist 1991), what conditions determine whether or not the quasar will have a jet? If crossing some threshold black hole mass constitutes a switch for a RLQ (Laor 2000), we should observe a mass boundary dividing the two AGN subgroups. Instead, we observe that RQQ form at all black hole and halo mass scales (Oshlack et al 2002; Woo & Urry 2002; McLure & Jarvis 2004). On average, however, RLQ decidedly have larger black hole masses (Metcalf & Magliocchetti 2006). While Laor (2000, figure 2) argues for a mass threshold for the radio loud/radio quiet dichotomy among the PG quasars, Sikora et al (2007, figure 4) find that division to be less sharp and Kelly et al (2008) report the existence of RQQ with black hole masses well beyond the alleged threshold value of about $10^9$ $M_\odot$. Although no mass boundary appears to exist, there seems to be a mechanism that increasingly favors the formation of a RLQ as the halo and black hole mass increase. As mentioned, whatever this mechanism is, it must also be compatible with recent observations of higher merger rates in RLQ. And, it must also allow for, or explain, the small fraction of RLQ compared to RQQ.

We show how to understand these observations in a context in which the formation of RQQ is preferred over RLQ for all black hole masses except at the higher mass end where the probability of forming the latter approaches 50%. Because the angular momentum of the black hole and galaxy should bear no correlation in the aftermath of a major merger, there is a chance that the conditions for radio loud quasar formation are realized in such systems. However, these objects, as we will describe, are unstable and require black holes that are massive compared to their accretion disks so a mass constraint further restricts their occurrence. Because spiral galaxies tend to feed their nuclei by secular processes in a less chaotic context, the conditions for RLQ formation are more restricted compared to elliptical galaxies that have recently experienced a merger. These ideas will shed light on a new understanding of what constitutes a selection effect on the conditions for RLQ formation. Despite a physical picture in which RLQ are at best equally likely to form as RQQ, exploring the radio loud/radio quiet dichotomy by using a range of black hole and halo masses, will introduce a bias giving the impression that radio loudness becomes dominant at the high mass end. In short, we show that while RLQ on average will have larger black hole masses than RQQ, the probability to form a RQQ is greater at all black hole masses. In Section II we quantitatively address the likelihood of forming both kinds of AGN and produce a theoretical plot of fractional probability as a function of black hole mass. From this we explain both the greater average black hole and halo mass for RLQ as well as the nature of the radio loud/radio quiet dichotomy. In other words, we show that the clustering nature of jetted versus non jetted quasars is fundamentally connected to the reason that jetted quasars constitute a minority among all AGN, thereby stringing together two seemingly disparate observations under one explanatory umbrella. In Section III we conclude.

II. Discussion

In the gap paradigm for black hole accretion and jet formation (Garofalo, Evans & Sambruna 2010), cold gas accretion around a high spinning black hole produces either a powerful jet (RLQ) or a negligible jet (RQQ) depending on the orientation of the disk. For counterrotation such as in Figure 1, a powerful jet coexists with a radiatively efficient disk, whereas for co-rotation the disk suppresses the jet (Figure 2). This orientation-based paradigm has been applied not only to the full gamut of observations across the entire AGN family, but in a scale invariant way also to stellar mass black holes. Most recently the model has allowed a new understanding of X-shaped radio galaxies (Garofalo, Joshi et al 2020), FR0 radio galaxies (Garofalo & Singh 2019), and neutron star X-ray binaries (Singh, Garofalo & Kennedy 2019).

According to King et al (2005) the fraction of accreting black holes that stably form retrograde configurations around the rotating black hole is determined by the angular momentum of the black hole $J_h$ and disk, $J_d$, via

$$f = \tfrac{1}{2}(1 - J_d/2J_h). \tag{1}$$

Because the total angular momenta of the black hole and disk are linearly related to the black hole and disk mass, respectively, Garofalo, Christian & Jones (2019) have translated condition (1) into a relative mass constraint and have explored the range of post-merger cold gas accreting mass ratios between disk and black hole. They argue that a larger range of mass ratios is possible in richer merger environments and translate that conclusion into a condition on the black hole mass. The bottom line is that most accretion disks are not stable in retrograde configurations and therefore flip to prograde ones unless the black hole mass is above a threshold value which is shown in Figure 3 (blue curve). At Log $M_{BH}/M_\odot$ = 8.5, where $M_{BH}$ is the mass of the black hole and $M_\odot$ is one solar mass, retrograde configurations are no longer excluded and become increasingly more likely. If the conditions for stability are reached, the probability of forming a retrograde disk becomes as likely as a prograde disk, and the blue curve therefore approaches a fractional probability of 0.5. Because the funneling of cold gas into the black hole will produce a quasar, the total probability of forming a retrograde and a prograde black hole adds to unity. Hence, the probability of forming a prograde disk above the mass threshold value decreases until it flattens out at 0.5 like the blue retrograde curve. Figure 3 amounts to a simplification of the stability condition, and that a more detailed analysis would likely produce red and blue curves whose first derivatives are not discontinuous as in Figure 3. Nonetheless, our qualitative conclusions remain unaffected.

The next step is to argue for an interpretation of Figure 3 as the fraction of RQQ and RLQ formed as a function of richness, with richness increasing with black hole mass. From this, note that above Log m = 9.25, the formation of RLQ is equal to that for RQQ. Hence, RLQ are not preferred over RQQ even in the richest environments (i.e. clusters). Second, at Log m < 9.25, i.e. in less rich environments (i.e. groups and fields), RQQ formation becomes increasingly preferred. Taken together, the simple overall conclusion to emerge from Figure 3 is that any average over the full range of possible richness factors for RLQ and RQQ will necessarily yield lower average richness factors for RQQ compared to RLQ.

Finally, note that it is possible to obtain the fraction of RLQ that are formed compared to the total quasar population from integrating over the range of masses to obtain a theoretical estimate of the fraction of RLQ relative to the total population. We can then compare to the 15%- 20% observed jetted AGN fraction. Notice that we have emphasized jetted AGN fraction and not jetted quasar fraction. This is due to the fact that in the model, jetted AGN that are not quasars, such as radiatively inefficient radio galaxies, have ancestors that were radio loud quasars. And Figure 3 includes them when they were in radiatively efficient mode. From Figure 3 we integrate the areas under the blue and red curves to obtain the total numbers of RLQ and RQQ and can therefore estimate the fraction of RLQ compared to the total population. We get 14. 2 percent.

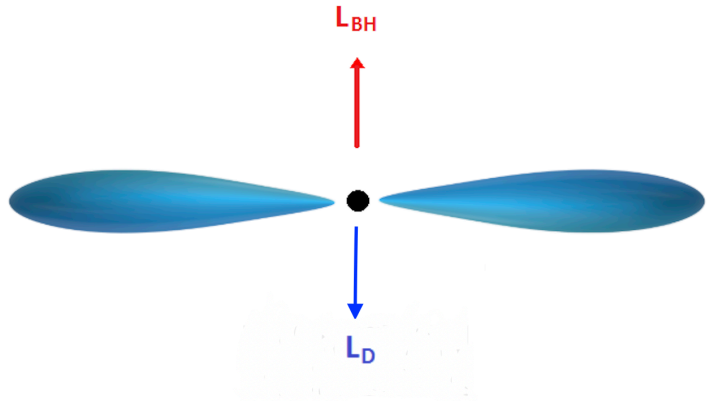

Figure 1: A counterrotating accretion disk with respect to the black hole with red showing the angular momentum direction of the black hole and blue showing the direction of the angular momentum of the accretion disk. Such a configuration has been shown to produce powerful, collimated, jets. These are radio loud quasars.

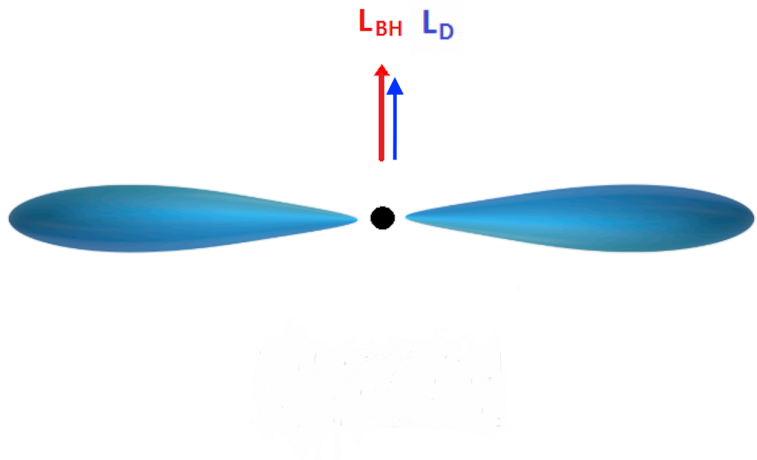

Figure 2: A corotating accretion disk with respect to the black hole with the colors as in Figure 1. Such a configuration has been shown not to produce powerful, collimated, jets. These are radio quiet quasars.

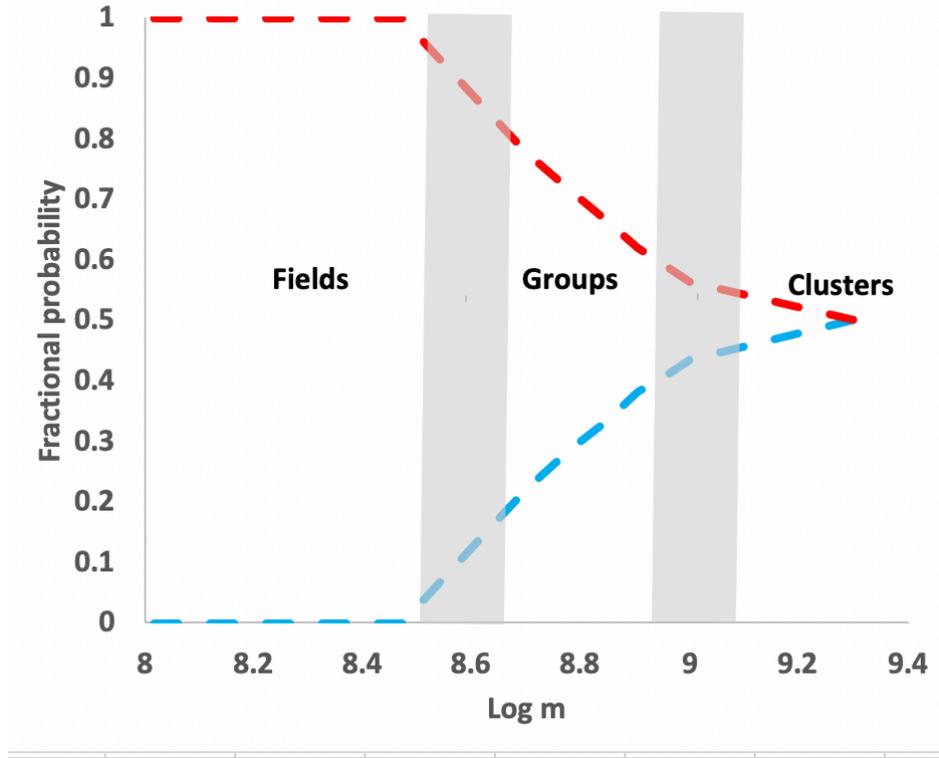

Figure 3: Fractional probability for RLQ (blue) and RQQ (red) versus log of black hole mass in solar masses, where m = $M_{BH}/M_\odot$ with the mass dependence of the fractional probability disappearing at the high mass end. As Log m increases, the blue curve approaches the red curve but always from below indicating that RLQ formation is never more likely than RQQ formation at any mass scale. Grey regions constitute rough boundaries for the expected locations of these black holes. Smallest black holes are expected to form in more isolated environments (fields), while the mass increases as one moves toward groups to richest environments (clusters).

III. Conclusions

We have explored the clustering of RQQ and RLQ from the perspective of a model that explains the radio loud/radio quiet dichotomy in terms of disk orientation around spinning black holes. Irrespective of the details of Figure 3, the fact that the RLQ formation curve (blue) approaches the RQQ formation curve (red) from below, explains why jetted quasars constitute a minority. We have been quantitative about this and obtained 14.16% for the jetted quasars compared to all quasars from Figure 3. The fact that the blue curve lives below the red curve allows us to explain both the minority of RLQ compared to RQQ, but also why the average black hole mass for RLQ is larger than that for RQQ. In short, we have been able to qualitatively extract explanations for both the apparent clustering nature of RLQ and RQQ as well as the reason why most AGN do not produce jets from the same theoretical constraint, namely disk orientation around spinning black holes. Although the time evolution of some radio loud quasars may lead to radio quiet quasars, these constitute a minority among the overall AGN phenomenon in the model and therefore the radio loud/radio quiet dichotomy can be understood without appealing to time evolution (see Garofalo et al 2016, and Garofalo, Christian & Jones 2019, for time evolution as connected to the time dependent merger function). The fact that radio loudness is not a dichotomy but a continuous phenomenon (Gurkan et al 2019) can also be understood in terms of a range in black hole spin values spanning the full retrograde/prograde regime. A prediction of all this is that as observations explore narrower high mass ranges of halo and black hole mass, the clustering preference of RLQ versus RQQ should decrease.

Acknowledgments

We thank the anonymous referee for brief but focused criticism.

References

1. Barnes, J.E. & Hernquist, L.E., 1991, ApJ, 370, L65
2. Chiaberge, M. et al 2015, ApJ, 806, 147
3. Garofalo, D, Evans, D.A., Sambruna, R.M. 2010, ApJ, 406, 975
4. Garofalo, D., Kim, M.I., Christian, D.J., Hollingworth, E., Lowery, A., Harmon, M., 2016, ApJ, 817, 2
5. Garofalo, D & Singh, C. B., 2019, ApJ, 871, 259
6. Garofalo, D., Christian, D.J., Jones, A.M., Universe, 5, 145
7. Garofalo, D., 2019, MNRAS, 489, 2
8. Gurkan, G. et al 2019, A&A, 622, A11
9. Hilbert, B. et al 2016, ApJS, 225, 12
10. Ivison, R.J. et al 2012, MNRAS, 425, 1320
11. Kelly, B.C., et al 2008, ApJS, 176, 355
12. King, A.R., Lubow, S.H., Ogilvie, G.I., Pringle, J.E., 2005, MNRAS, 363, 49
13. Laor, A. 2000, ApJ, 543, L111
14. McLure, R.J. & Jarvis, M.J., 2004, MNRAS, 353, L45

15. Metcalf, R.B. & Magliocchetti, M., 2006, MNRAS, 365, 101

16. Noirot, G. et al, 2018, ApJ, 859, 38

17. Oshlack, A.Y.K.N., Webster, R.L., Whiting, M.T., 2002, ApJ, 576, 81

18. Retana-Montenegro, E. & Rottgering, H.J.A., 2017, A&A, 600, A97

19. Sikora, M., Stawarz, L., Lasota, J.P., 2007, ApJ, 658, 815

20. Singh, C.B., Garofalo, D., Kennedy, K. 2019, ApJ, 887, 164

21. Woo, J.H. & Urry, C.M. 2002, ApJ, 581, L5

22. Worpel, H, et al 2013, ApJ, 772, 64

23. Wylezalek, D. et al 2013, ApJ, 769, 79

24. Zakamska, N.L. et al, 2019, MNRAS, 489, 497